\documentstyle[aps,twocolumn,epsf]{revtex}

\newcommand{\Od}{{\cal O}}

\newcommand{\qmixv}{(q_0,\vec{q},\tau)}
\newcommand{\qmixw}{(q_0,\omega_q,\tau)}

\newcommand{\tr}{\mbox{tr}}

\newcommand{\im}{\mbox{Im}}

\newcommand{\sgn}{\mbox{sgn}}

\newcommand{\fpi}{f_\pi}
\newcommand{\fpite}{f_\pi^t}
\newcommand{\fpisp}{f_\pi^s}
\newcommand{\ft}{f(t)}
\newcommand{\fpit}{f_\pi(t)}
\newcommand{\gpit}{g_\pi (t)}

\newcommand{\fpitsq}{f_\pi^2 (t)}
\newcommand{\ftsq}{f^2(t)}
\newcommand{\fzerosq}{f^2(0)}
\newcommand{\fdot}{\dot f(t)}
\newcommand{\fddot}{\ddot f(t)}

\newcommand{\tti}{\tilde t}

\newcommand{\lagf}{{\cal L}^{(4)}}

\newcommand{\intc}{\int_C dt \int d^3 \vec{x}}
\newcommand{\intcc}{\int_C d^4x}
\newcommand{\vxt}{(\vec{x},t)}

\newcommand{\be}{\begin{equation}}
\newcommand{\ee}{\end{equation}}
\newcommand{\ba}{\begin{eqnarray}}
\newcommand{\ea}{\end{eqnarray}}

\newcommand{\NP}[1]{{\it Nucl.\ Phys.\ }{\bf #1}}
\newcommand{\ZP}[1]{{\em Z.\ Phys.\ }{\bf #1}}

\newcommand{\PL}[1]{{\em Phys.\ Lett.\ }{\bf #1}}

\newcommand{\AN}[1]{{\em Ann. Phys. } (N.Y.) {\bf #1}}

\newcommand{\PRep}[1]{{\em Phys.\ Rep.\ }{\bf #1}}
\newcommand{\PR}[1]{{\em Phys.\ Rev.\ }{\bf #1}}

\newcommand{\IJmp}[1]{{\em Int.\ J.\ Mod.\ Phys.\ }{\bf #1}}

\def\IN{\relax{\rm I\kern-.18em N}}
\def\IR{\relax{\rm I\kern-.18em R}}
\def\ID{\relax{\rm I\kern-.18em D}}

\newcommand{\dnote}[1]{}
\begin{document}
%
%
\title{CHIRAL LAGRANGIANS OUT OF THERMAL EQUILIBRIUM
\thanks{Proceedings
of the 5th International Workshop on Thermal Field Theory,
Regensburg, Germany, August 10-14, 1998.}}

 
\author{Angel G\'omez Nicola\thanks{E-mail: 
gomez@eucmax.sim.ucm.es}}
\address{ Departamento de F\'{\i}sica 
Te\'orica\\ Universidad Complutense, 28040, Madrid, Spain}

\date{\today}



\maketitle


\begin{abstract}

We extend   chiral perturbation theory  to study 
 a meson  gas out of thermal equilibrium. We assume that  the system is  
initially in  thermal equilibrium  at a  temperature  $T_i<T_c$ and work 
within the   Schwinger-Keldysh contour technique. To lowest order and in the 
 chiral limit, we use  a nonlinear sigma model with a time-dependent 
 pion decay function $\fpit$, which we consistently define in terms of 
 the axial current two-point correlator. 
 As a first application, we analyse $\fpit$ up to 
 one loop order and  find  the $T_i$ dependence of the lowest  
 coefficients in its  short-time expansion. We discuss the applicability of 
 our results to  the time evolution of the  plasma formed after a heavy-ion 
 collision and discuss  the  interpretation of our model in a curved 
 space-time background.

\end{abstract}


\narrowtext

\section{Introduction}
\label{intro}
The chiral phase  transition plays a fundamental role in the description of   
 the plasma formed after a  relativistic 
 heavy-ion collision (RHIC). 
 For $N_f$ 
massless quarks (the chiral limit), QCD is invariant under the chiral 
 group $SU_L(N_f)\times SU_R(N_f)$. 
 That symmetry  is spontaneously broken at zero temperature to the vector 
  group $SU_V(N_f)$  (isospin), 
the  Nambu-Goldstone bosons (NGB) being  the lightest  
  mesons (pions for $N_f=2$ plus kaons and eta for $N_f=3$), which are 
    the relevant low-energy degrees of freedom. The light quark masses are 
 introduced perturbatively.  This picture 
 has proven to be very successful in describing   hadron observables 
 (like  decay constants or  scattering amplitudes),  for low   
 energies and small masses, through the 
 Goldstone theorem and current algebra. 
  On the other hand,   
 QCD at this scale  is strongly coupled and, therefore, if we want 
  to describe properly the meson dynamics we need  an effective  
 low-energy model. 
  Since $SU(2)\times SU(2)\rightarrow SU(2)$ is isomorphic to 
 $O(4)\rightarrow O(3)$  then for $N_f=2$ a candidate for the effective 
 theory 
  is the $O(4)$ linear sigma model (LSM), whose classical potential has the 
 familiar Mexican hat form.  Its  fundamental fields  are 
 the three pions and the $\sigma$, which acquires a nonzero 
 vacuum expectation value (VEV)  $v$.  However, one 
 drawback of the LSM is that its  coupling constant $\lambda$  becomes 
 large at  low energies, so that 
  perturbation theory in $\lambda$  is meaningless  and 
 alternative perturbative expansions must be considered, 
such as taking a large  number $N$ of NGB. 
 Besides, the LSM does no longer have  the QCD chiral symmetry breaking 
 pattern  for $N_f=3$. 
 An alternative approach is to construct an effective theory  as an 
 infinite sum of terms with increasing number of derivatives and with 
 all the QCD symmetries, only for the  NGB, 
 which are included in a  $SU(N_f)$-valued field, 
since $SU(N_f)$ is the coset space of the QCD 
 chiral symmetry breaking pattern.  To lowest 
 order, such  theory  is the so called 
 nonlinear sigma model (NLSM), which 
 allows, not only  to rederive  the current algebra  predictions 
 straightforwardly, 
but  to obtain  corrections to them in a consistent framework. Such 
 corrections   come both from  NGB loops and 
 higher order lagrangians and  can be  renormalised 
 order by order in energies, yielding unambiguous finite predictions
 for the observables. 
 This constitutes the so called chiral perturbation theory (ChPT) 
 \cite{we79,gale} 
 where  the  perturbative expansion is carried out in terms of the ratio of 
 the meson 
 energy scales of  the theory (masses, external momenta, temperature
 and so on) and 
  the  chiral scale $\Lambda_{\chi}\simeq$ 1 GeV. 
  For a review of low-energy QCD and effective lagrangians we refer to 
 \cite{dogoho92,dogolope97} and references therein.

 In thermal equilibrium at finite temperature $T$,  
the chiral symmetry is believed  to be restored at 
$T_c\simeq$ 150-200 MeV in a  phase transition, which is very 
 likely to be of  second order  
 for $N_f=2$ \cite{wi92}. It should be stressed 
 that  ChPT cannot strictly  establish the existence 
 of a phase transition since  it 
  provides  an  expansion in $T^2/\Lambda_{\chi}^2$ 
 for the order parameter $\forall T$. 
 Nonetheless, it predicts the correct behaviour of the observables 
as $T$ approaches $T_c$  (see below) and allows to study the meson gas  
 thermodynamics at low temperatures \cite{gale87,gele89}. To describe 
 the system near $T_c$,  the LSM seems  to be a better choice, since 
 it undergoes a second-order phase transition in the mean field 
approximation, $v(T)$ becoming  the order parameter. 
  However,  $T_c$  in the mean field LSM is independent of $\lambda$, 
 which  suggests that
  the NLSM should be equally valid to 
study the phase transition, which  is indeed the case provided one
 works  
 in the large $N$  limit  \cite{boka96}.

In the chiral limit and in 
 equilibrium at temperature $T$, the 
 only parameter of the NLSM lagrangian is a $T$-independent 
constant $f$ with dimensions 
 of energy, related  to the physical temperature-dependent 
 pion decay constant  as 
 $f_\pi (T)=f(1+\Od (p^2))$, 
 where $f_\pi\simeq 93$ MeV at $T=0$  
($\Lambda_\chi\simeq 4\pi\fpi$, which is the  typical loop  factor) 
 is   customarily 
 measured  in leptonic decays $\pi\rightarrow l\nu_l$. Here,   
$p$ denotes generically   any pion  momentum scale, 
 including   temperature  and  by  $\Od (p)$ we really  mean 
$\Od (p/\Lambda_\chi)$, as we explained before.   
Besides, 
 every pion loop  counts as $\Od(p^2)$ \cite{we79} and all the 
 infinities 
 coming from them can  be absorbed in the coefficients of 
 higher order lagrangians, which is also true for $T\neq 0$ \cite{gele89}. 
 The  next to leading order (NLO) one-loop 
 corrections induce the temperature dependence of 
 $\fpi (T)$    and for $N_f=2$  
in the chiral limit  the result is \cite{gale87,boka96}
\begin{equation}
\fpi^2 (T)=f^2\left(1-\frac{T^2}{6f^2}\right)
\label{fpieq}
\end{equation}

A few remarks are in order here. Firstly,  at $T\neq 0$,
due to the loss of Lorentz covariance in the thermal bath, 
 one should actually consider two independent pion decay constants,  
 $\fpite$ (temporal) and   $\fpisp$ (spatial),  which in general will 
 be  complex, their real and imaginary parts being  
 related respectively  with  the pion  velocity and damping rate  
 in the thermal bath  \cite{pity96}. 
  However, to NLO, one has  $\fpite=\fpisp=\fpi (T)$ in (\ref{fpieq}). 
 Notice that  (\ref{fpieq})  suggests a critical behaviour at 
 $T_c\simeq \sqrt 6 f_\pi$,  since  $\fpi (T)$  behaves roughly 
  like  $v (T)$, although they are different objects \cite{boka96}.
  Despite  (\ref{fpieq}) being just the lowest order  in the
 low temperature expansion (and thus not being  too accurate to extrapolate 
 it up to $T_c$) it predicts the right behaviour and a reasonable estimate 
 for $T_c$. 
  Another  important comment  is 
   that one has to be  careful
 when  defining $\fpi$ in a medium, trying 
 to extend naively low-energy theorems to $T\neq 0$ (see section \ref{fpi}).
 Nevertheless, 
there is a consistent way  of defining $f_\pi$ in the  thermal bath
from the axial-axial correlator spectral function
 \cite{boka96}  as it will be  explained in section \ref{fpi}.

The equilibrium assumption is not realistic if one is interested 
 in the dynamics of the expanding plasma formed after a RHIC, where there 
 are several observable non-equilibrium effects. One of them is the 
 formation of the so called disoriented chiral condensates (DCC), 
  regions in which the chiral field is correlated and has nonzero 
 components in the pion direction  \cite{an89}. 
 As the plasma expands, long-wavelength pion modes can develop 
 instabilities that grow very fast as the  field relaxes to the ground 
 state. If that  picture is correct, there would be observable effects, 
 such as coherent pion emission \cite{rawi93}. The question is whether those 
  domains can  grow fast and large enough to emit a significant number 
 of pions. A   necessary, but not sufficient, condition  is  that the 
 system is out of thermal equilibrium. This problem has been 
 extensively studied in the literature, using various 
non-perturbative and numerical 
approaches within  the LSM  
\cite{rawi93,bodeho95,bodeho97,giri95,cooper95,ladaco96}. 
 In all these 
 works, thermal equilibrium at $T_i>T_c$ is assumed for $t<0$  and then 
some mechanism  drives  the theory  off equilibrium for $t>0$,  
 encoded in the time dependence of the different lagrangian parameters.  
 The simplest choice  is  an external quench changing  
 the sign of the mass squared $m^2$  of the $O(4)$ field 
 so that modes with $k^2<m^2$ 
 become unstable
\cite{rawi93,bodeho95,bodeho97,giri95}. 
 A more realistic  approach is to describe  the plasma expansion in 
  proper  time and rapidity coordinates, which is analogous to study the 
 system in a certain space-time background metric, 
 assuming either boost  \cite{cooper95}  or scale   \cite{ladaco96}
 invariant  kinematics (cylindrical and spherical expansions, respectively). 
 During the time evolution, the
 pion modes  propagate as if they had  a negative mass squared (unstable 
 long-wavelength modes), 
 therefore  yielding exponential growth with time of  the pion propagator. 
 The general 
 conclusion  is that the typical size of the formed DCC regions is at most 
 3 fm, which appears to be 
not big enough to contain a large observable number 
 of pions, and  
the time during which the relevant plasma expansion takes place varies
 between  5-10 fm/c \cite{cooper95,ladaco96}.  
 Another  important nonequilibrium  effect  
is  the photon and dilepton production\cite{baier97,lebema97}. 
As these particles  are weakly interacting, they 
 decouple from the hot plasma carrying important information 
 about the dynamics,  so that  their final 
 spectra, dominated by meson decays, 
 is a promising signal to analyze nonequilibrium aspects \cite{baier97}.
 A recent suggestion is that a significant number 
 of those photons might have been produced 
 by $\pi^0$ decay during a typical DCC  nonequilibrium evolution 
 \cite{bodeho97}.

Our basic objective in this work will be 
 to construct an effective ChPT-based model to describe a meson 
 gas out of thermal equilibrium,  as an alternative to the  LSM 
 approach.    Our only degrees of freedom will 
  be then the  NGB  and we will consider  the most general 
 low-energy lagrangian  compatible with the QCD  symmetries.
 We will restrict here to  $N_f=2$  
and to the chiral limit. This is the simplest approximation we can consider 
 and  it allows  to build the model in terms of exact chiral symmetry.
 The lowest order lagrangian is then the NLSM.  
 In principle, we will be interested in the 
 physical regime  in which 
 the system is not far from equilibrium  (we will be more specific about 
 this point below), where an expansion in derivatives is consistent. 
 One of  the novelties of our approach is to exploit the
 analogy between near-equilibrium systems  and ChPT.

The structure of this work is the following: in section \ref{model} 
we introduce 
 our non-equilibrium NLSM and  establish the non-equilibrium ChPT,  
discussing how   the different symmetries are realised.
 We also discuss the interpretation of the model as a  
curved space-time QFT, paying special attention to renormalisation. 
The leading order non-equilibrium   pion propagator and 
 Lehman spectral function are also analysed  in that section, 
 which also contains  most of 
 our notation and  conventions. 
 In section \ref{nloprop}, we derive  the one-loop  NLO correction to the 
 propagator, which will be needed later.  
 In section  \ref{fpi}  we provide a consistent  definition of the  
 nonequilibrium time-dependent  pion decay  
 functions (PDF) extending the equilibrium pion decay constants 
 and  we derive them up  to one loop in ChPT. 
  One of the motivations to concentrate on  $\fpit$ first is that 
 to one loop in ChPT we expect that all physical observables are obtained 
 from the tree level ones by replacing $\ft\rightarrow \fpit$, as it 
 happens indeed in equilibrium. 
  We analyse  the short-time behaviour of  $\fpit$ and estimate the 
relevant time scales  involved, discussing the  role of unstable modes.
 The issue of  axial current conservation is   
 also analysed in section \ref{fpi}. 
Finally, in section \ref{conc},  we present our conclusions and 
discuss some open questions, as well as  further applications  of our 
 approach.

\section{The NLSM and ChPT out of equilibrium}
\label{model}

\begin{figure}\epsfxsize=10cm
\vspace*{-5.5cm}
\centerline{\epsfbox{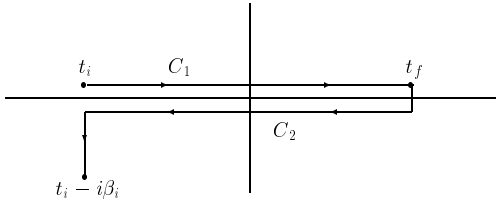}}
\vspace*{-5.5cm}
\vskip 4mm
\caption {The contour $C$ in complex time $t$. The 
 lines $C_1$ and $C_2$ run  between $t_i+i\epsilon$ and 
 $t_f+i\epsilon$ and  $t_f-i\epsilon$ and $t_i-i\epsilon$ 
respectively, 
 with  $\epsilon\rightarrow 0^+$.}
\label{fig1}
\end{figure}

 We will  assume that the system is  in thermal equilibrium for $t<0$ at a 
 temperature $T_i<T_c$ and  for $t>0$ we let the  lagrangian parameters 
 be  time-dependent (we are also   assuming that the 
 system is homogeneous 
 and isotropic).  The generating functional 
  of the theory can then be formulated in the  path integral formalism, 
  by letting the  time 
 integrals run over the Schwinger-Keldysh 
contour  $C$ showed in Fig.\ref{fig1}  
\cite{schkel60s,sewe85,chin85,boleesi93}. 
 We will eventually let  $t_i\rightarrow -\infty$ and 
 $t_f\rightarrow +\infty$, although we will show below that 
our results are independent of $t_i$ and $t_f$. 
  We remark that, even in that limit, 
  the imaginary-time leg of the contour has to be kept, since it 
encodes the KMS equilibrium boundary conditions \cite{sewe85,boleesi93}.

With the above assumptions, our low-energy model will be the 
 NLSM with $f$ becoming  a real function $f(t)$, 
\begin{equation}
S[U]=\intcc \ \frac{f^2(t)}{4} \ \tr \ \partial_\mu U^\dagger\vxt
\partial^\mu U\vxt
\label{nlsmne}
\end{equation}
with  $\intcc\equiv\intc$, $f(t<0)=f$  and $U\vxt\in SU(2)$ is 
the NGB field, satisfying 
 $U (\vec{x},t_i+i\beta_i)=U(\vec{x},t_i)$ with $\beta_i=T_i^{-1}$. 
 We will parametrise $U$  as
\ba
U\vxt=\frac{1}{f(t)}\left\{\left[f^2(t)-\pi^2 \right]^{1/2} I
+i\tau_a\pi^a\right\}
\label{upar}
\ea
where $\pi^2=\pi^a\pi_a$,  $\pi^a\vxt$ the pion fields satisfying 
 $\pi^a (t_i+i\beta_i)=\pi^a (t_i)$ ($a=1,2,3$) and   
$I$ and $\tau_a$ are the identity and Pauli matrices. 
In terms of the  $\pi^a$ fields, the action 
(\ref{nlsmne}) becomes 
\begin{eqnarray}
S[\pi]&=&\intcc \frac{1}{2}\left\{ \partial_\mu \pi^a \partial^\mu \pi^b 
\left[ \delta_{ab}+\frac{\pi_a \pi_b}{f^2(t)-\pi^2}\right]\right.
\nonumber\\
&+&\left.\frac{f(t)\fdot}{\ftsq-\pi^2}
\left[\frac{\fdot}{\ft}\pi^2-2\pi^a\dot\pi^a\right]\right\}
\label{nlsmnepi}
\end{eqnarray}
 
 Note that by  including  
 $\ft$ in the parametrisation of the NGB field  we recover the 
 canonical kinetic term in the action. Other choices  amount to a 
 time-dependent normalisation of the pion fields, which  should not have 
 any effect on the physical observables  (see section \ref{fpi}). 
 For instance, we could redefine 
 $\tilde\pi^a=\pi^a f(0)/\ft$, so that 
  for the $\tilde\pi^a\vxt$ fields the action is 
  the  equilibrium  NLSM multiplied by the  time-dependent scale 
factor $\ftsq/\fzerosq$, which is directly interpreted in terms of 
 a curved space-time background, as we will see below.


The action (\ref{nlsmne}) is manifestly chiral invariant, i.e, under   
 \mbox{$U\rightarrow LUR^\dagger$} with $L$ and 
 $R$  arbitrary constant 
$SU(2)$ matrices. Notice that, as we are working in the 
 chiral limit by assumption, we have not considered a pion 
mass term in the 
 action. Such  term would break  explicitly the chiral symmetry, 
 being still  
invariant under vector isospin $L=R$ transformations for  $m_u=m_d$. 
  The conserved vector $V_\mu^a$ and 
 axial-vector $A_\mu^a$  Noether currents for the chiral symmetry can be 
 derived  by gauging the symmetry  and then taking 
 functional derivatives with respect to the external sources 
 \cite{gale}. Extending that procedure   to our non-equilibrium
 action  (\ref{nlsmne}) we get
\begin{eqnarray}
A_\mu^a\vxt&=&i\frac{\ftsq}{4}\tr\left[\tau^a\left(
U^\dagger\partial_\mu U-U\partial_\mu U^\dagger\right)\right]
\nonumber\\
V_\mu^a\vxt&=&-i\frac{\ftsq}{4}\tr\left[\tau^a\left(
U^\dagger\partial_\mu U+U\partial_\mu U^\dagger\right)\right]
\label{currents}
\end{eqnarray}

 The model is also parity invariant. Lorentz 
 covariance is lost, because we have chosen the frame in which  the 
 thermal bath is at rest for $t<0$. The  time translation  symmetry is  also 
 broken at nonequilibrium.

Let us now discuss how to implement a consistent ChPT out of 
 equilibrium. The new ingredient  with respect to 
 the equilibrium case, is the temporal variation of $\ft$. We will then
 consider 
\begin{equation}
\frac{\fdot}{f^2 (t)}\simeq \Od(p), \qquad 
\frac{\fddot}{f^3(t)}, \frac{[\fdot]^2}{f^4(t)}\simeq\Od (p^2), 
\label{chipoco}  
\end{equation}
 and so on, 
  the rest of the chiral power counting being the same as in equilibrium, 
  i.e, $T_i=\Od (p)$, $\partial\pi=\Od(p^2)$ and every pion loop is 
 $\Od(p^2)$. 
  Therefore, in our approach we treat  the deviations of 
 the system from equilibrium perturbatively, following  the ChPT guidelines. 
 Thus, with this  chiral power counting, 
we can  expand our action 
(\ref{nlsmne}) 
 to the relevant order in pion fields and calculate any 
observable by taking 
 into account all the Feynman diagrams that contribute to that order. 
The loop  divergences should be such that they can be absorbed 
 in the coefficients of  higher order lagrangians, 
which in general will require 
the introduction of new time-dependent 
  counter-terms (see below).

 Notice that according to  (\ref{chipoco}), we can always 
 describe the short-time non-equilibrium regime, just by 
 Taylor expand $\ft$ around $t=0$. In fact, 
for times $t\leq \fpi^{-1}$, 
 the Taylor expansion of $\ft$ is equivalent to a chiral expansion,
 since then $\dot f (0) t/f=\Od (p)$, $\ddot f(0)t^2/f=\Od (p^2)$ and so on. 
 In case we are interested only in  short times,  the specific 
 form of $f(t)$ will  not be  important (see section \ref{fpi}),  
 but we want to stress that the 
  conditions (\ref{chipoco}) do not constrain  us to work  at short 
 times, but just to remain  close enough to equilibrium.


 To leading order in the pion fields, the action (\ref{nlsmnepi}) reads
\begin{eqnarray}
S_0[\pi]&=&-\intcc \frac{1}{2}\pi^a\vxt\left[  \Box + 
m^2 (t)\right] 
\pi^a\vxt \label{actlo}  \\
m^2(t)&=&-\frac{\fddot}{\ft}
\label{msqt}
\end{eqnarray}
where we have integrated by parts, using  the boundary 
conditions for the pion fields. 
 Thus, we see that the leading order non-equilibrium effect of our model can 
 be written as  
 a time-dependent mass term  for the pions. 
As we commented before, time-dependent  field masses are  a  common 
 (and welcomed) 
   feature of out of equilibrium models  
\cite{bodeho95,bodeho97,giri95,cooper95,ladaco96}.
 In our  approach, that mass parametrises the deviation of the 
 plasma from equilibrium through the temporal variation of $\ft$. 
 Notice that $m^2(t)$ can be 
 negative, so that our model accommodates  unstable pion 
modes, whose importance we have discussed before. 
 The effect of those modes should not be important for the short-time 
 evolution, but for longer times we expect them to influence significantly 
 the dynamics.

 We remark that  
 the axial current is classically   conserved  despite the 
  existence of the time-dependent  mass term. For instance, 
to leading order  we have, from  (\ref{currents}), 
 \be
A_\mu^a=-\ft\partial_\mu\pi^a+\delta_{\mu 0}\fdot\pi^a +\Od (\pi^3)
\quad ,
\label{axcurrlo}
\ee
which satisfies $\partial^\mu A_\mu=0$ using the  
 equations of motion to the same order  
$\left[\Box + m^2 (t)\right] \pi^a=0$.   
 Had we included an explicit 
 pion mass term $m_\pi\simeq$ 140 MeV, the axial current 
 would not have been conserved and the effective mass term in the 
 leading order  action 
 would have been proportional to $m_\pi^2+m^2(t)$,  
 so that the onset for instabilities would be rather 
 $m^2(t)<-m_\pi^2$.

 We will now  rephrase our model in the language of  
 curved space-time background QFT \cite{bida82}, which will turn out to 
 be a very  useful analogy. 
For that purpose, let us   consider the NLSM  in an arbitrary background
 space-time metric $g_{\mu\nu}$ \cite{dole91}, 
\begin{eqnarray}
S_g[\tilde\pi]&=&\intcc \frac{1}{2} \sqrt{-g} g^{\mu\nu}    
\left\{ \partial_\mu \tilde\pi^a \partial_\nu \tilde\pi^b 
\left( \delta_{ab}
\right.\right.\nonumber\\
&+&\left.\left.\frac{\tilde\pi_a \tilde\pi_b}{f^2(0)-\tilde\pi^2}
\right)\right\}+\xi S_R[\tilde\pi,R] \quad ,
\label{nlsmcurv}
\end{eqnarray}
plus $\tilde\pi$ independent terms, where $g$ is the metric determinant and 
the  term 
$\xi S_R[\tilde\pi,R]$ accounts for the possible  
couplings between the  pion fields and the 
 scalar curvature $R (x)$ (like  $R(x)\phi^2$  for a free scalar field 
$\phi$  \cite{bida82}). 
 To compare with our nonequilibrium model, we will choose a 
 spatially flat Robertson-Walker (RW)  space-time:
\begin{equation}
 ds^2=dt^2-a^2(t)d\vec{x}^2
\label{RW}
\end{equation}
with  $a(t)$  the scale factor. For $t<0$ we choose 
 the space-time as Minkowskian, so that  $a(t<0)=1$ and the system  in 
 thermal equilibrium at temperature $T_i$.
  Then, if we change  coordinates 
 to the conformal time $\eta$ defined by $t=\int_0^\eta a(\eta')d\eta'$, 
  the  metric becomes $ds^2=a^2(\eta)[d\eta^2-d\vec{x}^2]$ and we see that the 
 action in (\ref{nlsmcurv}) for the
  minimal coupling case 
 $\xi=0$ is nothing but $S[U]$ 
 in (\ref{nlsmne}), using the $\tilde\pi$ parametrisation discussed before 
 and  identifying $a(t)=f(t)/f(0)$. 
 That is, our non-equilibrium model in the chiral limit  is 
 equivalent to a 
 RW spatially flat space-time in conformal time, 
that starts expanding (or contracting) 
 at $t=0$ with a rate given by  the  scale factor $f(t)/f(0)$ and couples 
 minimally to the matter fields. Our model is then not only suitable as 
 an effective theory for a RHIC environment, but also in a cosmological 
 framework. Nonequilibrium models for RHIC in which the plasma expansion 
 is also parametrised in a background metric have been analysed 
 in \cite{cooper95,ladaco96} (see comments in section \ref{intro}).  

Let us now consider  $\xi\neq 0$. The lowest order  $S_R$ 
term we can construct has the form of an effective mass term, 
\be
S_R[\tilde\pi,R]=R(t) f^2 (0)
\left[1-\frac{\pi^2}{f^2(0)}\right]^{1/2}\quad ,
\label{xiterm}
\ee
 so that it breaks explicitly 
the chiral symmetry.  Let us then  pick up 
the lowest order contribution in pion fields  
 $S_R=-R\tilde\pi^2/2$ in (\ref{xiterm}).  Then recalling  the 
 expression of the scalar curvature for the spatially flat RW metric 
in conformal time
 $R(\eta)=6 \ddot f(\eta)/f^3(\eta)$, we see  that the curvature term 
  exactly cancels  the $m^2(t)$ term in (\ref{actlo}), after rescaling 
  $\tilde\pi(x)=f(0) \pi(x)/f(t)$, provided 
 we choose $\xi=1/6$. But that value 
 of $\xi$ is precisely the only one making  the theory scale invariant 
\cite{bida82}, so that in that case the  
 leading order lagrangian 
would be scale invariant but not chiral invariant. Thus,  
the lagrangian chiral and conformal 
 symmetries  are not  compatible in a curved background \cite{dole91} or,
 equivalently, at  nonequilibrium.  Thus, 
 for $\xi=0$ we interpret the mass term $m^2(t)$ 
as the minimal coupling with the background, which for  $m_\pi=0$
 yields chiral invariance of the 
 lagrangian, though 
  breaking the  conformal symmetry. This is the choice we will  
 adopt here, since we want to preserve chiral symmetry. 

 A very important result is that 
  all the one-loop divergences arising 
 from  (\ref{nlsmcurv}) can be absorbed 
 in the coefficients of the $\Od (p^4)$ lagrangian $\lagf$, which 
 includes new chiral-invariant 
couplings of pion fields  with the  curvature \cite{dole91}. 
In the chiral limit, for instance, there are two such new
 terms and  therefore two more undetermined constants 
in addition to the Minkowski case ones (see \cite{dole91} for details). 
  Therefore, following this analogy, 
we know how to  construct the higher order 
 nonequilibrium lagrangians that eventually will absorb all the divergences 
 we may find in our analysis. We will come back to this point below.

From the generating functional
 we can derive in the standard manner, the Green functions 
 ordered in time along the contour $C$ \cite{boleesi93}. 
 The two-point 
 function defines the pion propagator
\begin{equation}
G^{ab} (x,y)=-i < T_C \pi^a (x) \pi^a (y) >
\label{propdef}
\end{equation}
with $T_C$ the time-ordered product along $C$. 
 To leading order the   pion propagator is 
$G_0^{ab} (x,y)=\delta^{ab}G_0(x,y)$ (by isospin invariance), 
with $G_0(x,y)$   the  solution of the differential equation 
\begin{equation}
\left\{\Box_x + m^2(x^0)\right\} G_0 (x,y)=-\delta^{(4)} (x-y)
\label{loprop}
\end{equation}

As for the boundary conditions, 
 firstly we have to impose  KMS equilibrium  conditions on  
 the imaginary leg of the contour, that is, 
\begin{equation}
G^>_0 (\vec{x},t_i-i\beta_i;y)=G^<_0 (\vec{x},t_i;y)
\label{kms}
\end{equation}
with the advanced and retarded propagators  defined as customarily along the 
 contour $C$.
In addition,  the propagator  must be  
 continuous and differentiable $\forall t\in C$, so that 
 the solution is uniquely defined   \cite{sewe85}. 
  Therefore, in our case, we will demand  
 that the solution of (\ref{loprop}) for $t\rightarrow 0^+$ matches 
 that for $t<0$, which is the equilibrium solution, given 
 in momentum space by \cite{lebe96}:
\be
G_0^{eq} (q_0,\omega_q)= \rho^{eq}_0 (q_0,\omega_q) 
 \left[\theta_C(t-t')+n_B(q_0)\right]
\label{g0eq}
\ee
 where $w_q^2= |\vec{q}|^2$, \mbox{$n_B(x)=[\exp(\beta x)-1]^{-1}$} is the 
 Bose-Einstein distribution function, 
$\theta_C$ is the step function along the  contour $C$ and  
 \mbox{$\rho_0^{eq}=-2\pi i \sgn (q_0) \delta (q_0^2-\omega_q^2)$}.

We will make use of two different representations for the non-equilibrium 
Green functions. Notice 
 that $G(x,x')=G(t,t',\vec{x}-\vec{x'})$ 
 because of the  lack of time translation invariance. 
Therefore, we will  define, as 
 customarily,  the ``fast'' temporal variable $t-t'$ and the ``slow'' one 
 $\tau\equiv (t+t')/2$,   so that 
$F(q_0,\omega_q,\tau)$ and $F(\omega_q,t,t')$  will denote, respectively, the 
 fast and mixed (in which only the spatial coordinates are transformed) 
 Fourier transforms of $F(x,x')$. The fast Fourier transform depends 
 separately  on  $q_0$ and $\omega_q$ because of the loss of 
 Lorentz covariance and has the extra nonequilibrium $\tau$-dependence.



 In the mixed representation, (\ref{loprop}) becomes
\be
\left[\frac{d^2}{dt^2}+\omega_q^2+m^2(t)\right] G_0 (\omega_q,t,t')
=-\delta (t-t')
\label{lopropmix}
\ee

The general solution of (\ref{lopropmix}) cannot be found analytically for 
 $m^2(t)$ arbitrary, but it can formally be written as a 
 Schwinger-Dyson integral equation in terms of the equilibrium solution 
 as 
\ba
\lefteqn{G_0(\omega_q,t,t')=G_0^{eq} (\omega_q, t-t')}
\nonumber\\&+&
\int_C dz m^2(z) G_0^{eq} (\omega_q, t-z) G_0(\omega_q,z,t') 
\label{sdg0eq}
\ea
with $G_0^{eq} (\omega_q, t-t')$ the solution of (\ref{lopropmix}) with 
 $m^2(t)=0$.

Another object of interest for our purposes is the Lehman spectral 
function, defined  as \cite{chin85,lebe96}
\begin{equation}
\rho (x,y)=G^> (x,y)-G^< (x,y)
\label{rho}
\end{equation}

Recall that in equilibrium and to leading order, the spectral function is 
$\rho_0^{eq}$ in (\ref{g0eq}). Let us now discuss 
 some useful properties of the spectral function. Firstly,  by definition 
 $G(x,y)=G(y,x)$ and hence $G^>(x,y)=G^<(y,x)$. Therefore, 
\mbox{$\rho(x,y)=-\rho(y,x)$} and  
\mbox{$\rho (q_0,\omega_q,\tau)=-\rho(-q_0,\omega_q,\tau)$}.
The normalisation of $\rho_0$ is the same as in equilibrium, i.e,  
 for  the  parametrisation (\ref{upar}) we have
\begin{equation}
\frac{1}{2\pi i}\int_{-\infty}^{+\infty} q_0 \rho_0(q_0,\omega_q,\tau)=
 \left.\frac{d \rho_0 (\omega_q,t,t')}{dt}\right\vert_{t=t'}=
 -1
\label{rho0norm}
\end{equation}

 It is not difficult to check (\ref{rho0norm}), by using for instance 
(\ref{sdg0eq}) and  $\rho_0^{eq} (\omega_q,t,t)=\rho_0 
 (\omega_q,t,t)=0$. 


\section{Next to leading order propagator}
\label{nloprop}

 We will now  obtain the NLO correction to the two-point 
 Green function (\ref{propdef}). 
 For that purpose, we need the action in 
 (\ref{nlsmnepi}) up to four-pion terms: 
\begin{eqnarray}
S[\pi]&=&S_0[\pi]+\frac{1}{2}\intcc
\left\{\frac{1}{\ftsq}
\left[ \partial_\mu\pi^a\partial^\mu\pi^b\pi_a\pi_b
\right.\right.\nonumber\\
&+&\left.\left.
\frac{1}{2}(\pi^2)^2\left(\frac{\fddot}{\ft}
-\frac{\dot f^2(t)}{\ftsq}\right)
\right] +\Od (\pi^6)\right\}
\label{nlsmnenlo}
\end{eqnarray}

\begin{figure}\epsfxsize=12cm
\vspace*{-6.5cm}
\centerline{\epsfbox{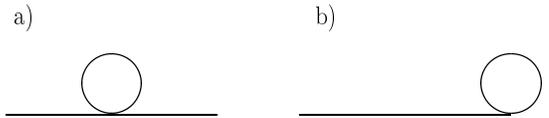}}
\vspace*{-8cm}
\vskip 4mm
\caption {One-loop diagrams contributing to 
 a) The NLO pion propagator and b) The NLO axial-axial correlator}
\label{fig2}
\end{figure}

The NLO correction to the two-point function is given by 
 the tadpole diagram a) in Fig.\ref{fig2}. It is important to bear in mind
 that in principle 
 we should also include a tree diagram coming from the $\lagf$ 
lagrangian, which would take care of renormalisation. 
We shall ignore that contribution here and we will justify this approximation 
 in section \ref{fpi} in the context of the short-time regime.

After summing  over  
all isospin indices in the loop, we can  use 
 the differential equation  (\ref{loprop}).  
 However, in doing so we  have to deal with the 
 divergent quantity  $\delta^{(4)}(0)$.  We will use 
 dimensional regularisation, which is a suitable 
 scheme to deal with ChPT \cite{dogolope97}, 
so that  $\delta^{(d)}(0)=0$.  Therefore, 
 we can  replace 
 $\Box_z G_0(z,z)=-m^2(z^0)G_0(z,z)$. 
Let us  concentrate on $t,t'\in C_1$ in Fig.\ref{fig1} 
(i.e, the $G_{11}$ component of the propagator \cite{lebe96})
 and  take $t$ and $t'$ positive, 
which is actually the relevant case for our purposes, 
 since the nonequilibrium effects start for positive times.
 Then,  in the mixed representation  we obtain
\begin{eqnarray}
\lefteqn{G_{NLO,11}^> (t,t')=G_{0,11}^> (t,t')}
\nonumber\\
 &-&\frac{1}{2\omega_q}
\coth\left[\frac{\beta_i\omega_q}{2}\right] \frac{G_0^{eq}}{f^2(0)}
\cos\left[\omega_q (t+t')\right]\nonumber\\
&+&i\left\{\int_{0}^t d\tti \left[\Delta_1(\tti,\omega_q)G_0^> (\tti,t)
G_0^>(\tti,t')\right.\right.\nonumber\\
&+&\left.\Delta_2(\tti)\dot G_0^> (\tti,t) \dot G_0^>(\tti,t')\right]
\nonumber\\ 
&-&\int_{0}^{t'} d\tti \left[\Delta_1(\tti,\omega_q)G_0^< (\tti,t)
G_0^<(\tti,t') \right.\nonumber\\&+& 
\left.\Delta_2(\tti)\dot G_0^< (\tti,t) \dot G_0^<(\tti,t')\right]
\nonumber\\
&-&\int_{t'}^t d\tti \left[\Delta_1(\tti,\omega_q)G_0^< (\tti,t)
G_0^>(\tti,t')\right.\nonumber\\
&+&\left.\left.
\Delta_2(\tti)\dot G_0^< (\tti,t) \dot G_0^>(\tti,t')\right]\right\}
\label{nloposit}
\end{eqnarray}
and $G_{NLO,11}^< (t,t')=G_{NLO,11}^> (t',t)$, where 
we have suppressed for simplicity the $\omega_q$ dependence of 
 the propagators,  the dot denotes $d/d\tti$,


\ba 
\lefteqn{
\Delta_1 (\tti,\omega_q)=\frac{1}{f^2(\tti)}\left[\left(
6\frac{\ddot f(\tti)}{f(\tti)}-
5\left(\frac{\dot f(\tti)}{f(\tti)}\right)^2\right.\right.}
\nonumber\\
&-&\left.\left.\omega_q^2
\right) G_0(\tti)
-2\ddot G_0(\tti)+4\frac{\dot f(\tti)}{f(\tti)}\dot G_0(\tti)
\right]\quad ,\label{del1}\\
\Delta_2(\tti)&=&\frac{G_0(\tti)}{f^2(\tti)}\quad ,
\label{del2}
\ea
 $G_0(z^0)\equiv G_0(z,z)$ is the time-dependent 
equal-time correlation function and $G_0^{eq}=G_0(\tti<0)$, which   
is $\tti$ independent and is given in dimensional regularisation by
\begin{eqnarray}
G_0^{eq}=-i\int \frac{d^{d-1}q}{(2\pi)^{d-1}}\frac{1}{2\omega_q} 
 \coth\left[\frac{\beta_i\omega_q}{2}\right]=-i\frac{T_i^2}{12}
\label{g0eqdr}
\end{eqnarray}
 
The equilibrium result is finite in the 
 chiral limit, where 
there  is no need to renormalise  the tadpole diagrams in 
 dimensional regularisation \cite{dogoho92,dogolope97}. However, out of 
 equilibrium, $G_0(t)$ is in general a divergent time-dependent 
quantity, even in the  chiral limit, which means that the infinities have 
 to be absorbed in the new nonequilibrium counterterms we discussed in 
 section \ref{model}.

  To derive (\ref{nloposit}) we have used the equilibrium result 
(\ref{g0eq}) to evaluate explicitly the $\tti<0$ integrals.  
 We observe that (\ref{nloposit}) is $t_i$ 
and $t_f$ independent, which is a good check of consistency of our 
 result. It can be checked that the result is 
also $t_i$ and $t_f$ independent for 
 $G_{12}$, $G_{21}$ and  $G_{22}$. Another consistency check is obtained 
 by replacing the equilibrium $G_0(\omega_q,t,t')$ and 
$G_0(t)\rightarrow G_0^{eq}$ in 
 (\ref{nloposit}), so that we recover the equilibrium result
\begin{equation}
G_{NLO,11}^{eq>} (t-t')=G_{0,11}^{eq>} (t-t')
\left(1-\frac{T^2}{12f^2}\right)
\label{nlogeq}
\end{equation}

The  above result agrees with   \cite{boka96}. Notice
 that, in turn,  we have derived the correct equilibrium result using the  
 general  contour $C$, which includes  both imaginary-time and real-time 
 thermal field theory.

\section{The nonequilibrium pion decay functions}
\label{fpi}

 At $T=0$ the pion decay constant  is customarily defined through 
 the PCAC low-energy theorem for the 
  expectation value of the  axial current between 
 the vacuum and an asymptotic pion state  and 
 one  relates that amplitude 
 with Green functions via the LSZ formula \cite{dogoho92}. 
 However, in a thermal bath, the 
  concepts of LSZ and asymptotic states are subtle and it is much more 
 convenient to work directly with thermal 
Green functions. 
The  object of interest for our purposes  is  the axial-axial 
 correlator 
\be
A_{\mu\nu}^{ab} (x,y)=<T_C A_\mu^a (x) A_\mu^b (y) >
\label{axax}
\ee

At $T=0$, the longitudinal part of the axial-axial 
 spectral function  is proportional, in the chiral limit, 
 to a \mbox{$\delta$-function}, 
 signaling  the  existence of massless  NGB (see 
 below) and the 
 proportionality constant is  $\fpi$. 
 However, at $T\neq 0$ that definition is more subtle: firstly, 
because of the 
  loss of covariance, the 
 tensorial structure compatible with current conservation 
is more general than at $T=0$ \cite{kash94} and 
one can define  two  independent and complex $\fpisp$ (spatial) 
 and $\fpite$ (temporal) \cite{pity96}. Secondly, the 
pion dispersion relation
 is not in general a $\delta$-function as in vacuum. Nevertheless, a proper 
 definition of $\fpi (T)$ can still 
be given \cite{boka96}, which, to one-loop 
 gives the result (\ref{fpieq}) with $\fpisp=\fpite=\fpi$  real. To higher 
 orders, $\fpisp\neq\fpite$ and $\im \fpisp$, $\im \fpite$ are nonzero
 \cite{pity96}.



Let us then  analyze the axial-axial correlator (\ref{axax}) in our 
 non-equilibrium model. As we said above, the relevant quantity, as far as 
 $\fpi$ is concerned, is not 
  (\ref{axax}), but its spectral function $\rho^{ab}_{\mu\nu} (x,y)$, 
defined by subtracting its advanced 
 and retarded parts. Again, we write 
 $\rho^{ab}_{\mu\nu}=\delta^{ab}\rho_{\mu\nu}$ by isospin invariance and 
 we concentrate on the  fast  Fourier transform 
$\rho_{\mu\nu} \qmixv$.  Firstly, we will  
write down the most general form for 
 $\rho_{\mu\nu}$ compatible with the symmetries 
 and analyse the restrictions imposed   by the axial current conservation
 Ward Identity (WI). 

 From the definition of the 
 spectral function, we readily realise that 
$\rho_{\mu\nu} \qmixv=-\rho_{\nu\mu} (-q_0,-\vec{q},\tau)$. Then, 
 from rotational invariance, 
\be
\rho_{ij} \qmixv = q_i q_j \rho_L \qmixw + \delta_{ij} \rho_d \qmixw
\label{rhoij}
\ee
with $\rho_{L,d}(q_0)=-\rho_{L,d}(-q_0)$ 
\footnote{$\rho_L$ and $\rho_d$ correspond in equilibrium and 
in the notation of \cite{boka96},  to 
 $\sgn(q^0)\rho_A^L q_0^2/\omega_q^2 q^2$ and 
$\sgn(q^0)\rho_A^T$ respectively.}. On the 
other hand,  $\rho_{jo}$ is  a 3-vector, so that  
\be
\rho_{j0} \qmixv = q_j \rho_S \qmixw
\label{rhoj0}
\ee
and $\rho_{j0}\qmixv =q_j \rho_S (-q_0,\omega_q,\tau)$. 
Thus, $\rho_{\mu\nu}$ is characterized, in principle by the four 
  functions $\rho_L$, $\rho_d$, $\rho_S$ and $\rho_{00}$. 
 However, not all of them are independent. 
We have seen  that the axial current is 
 classically conserved in our model and,  therefore,  the WI   
 $\partial^x_\mu \rho_{\mu\nu} (x,y)=\partial^y_\nu 
\rho_{\mu\nu} (x,y)=0$ 
  must hold (there are no axial flavour anomalies for $N_f=2$). 
 It can be checked, following standard path integral 
 methods \cite{dogolope97}, 
 that the same equilibrium  WI holds out of equilibrium.
  Using the symmetry properties discussed before,  the WI yields
\ba
q^0 \rho_{00} -\omega_q^2 \rho_S  -\frac{i}{2}\dot\rho_{00} &=&0
\nonumber\\
q^0 \rho_S  - \omega_q^2 \rho_L +\frac{i}{2}\dot\rho_S 
+ \rho_d &=&0
\label{wirho}
\ea
where the dot denotes $\partial/\partial\tau$. Thus, only two components 
 of $\rho_{\mu\nu}$ 
are independent, as in equilibrium \cite{kash94,boka96}. 
 Notice that the time derivatives  are not present in  equilibrium.

At $T=0$ one 
 has $\rho_L=2\pi \fpi^2\sgn(q^0)\delta(q^2)$, which defines $\fpi$ 
 and states that in the chiral limit there are 
 NGB with the same quantum numbers as the axial 
 current, according to Goldstone theorem. That is not the case 
 in equilibrium at $T\neq 0$ \cite{kash94,boka96}, where to define 
 properly $\fpi (T)$ requires taking  the $\omega_q\rightarrow 0^+$ limit, 
 in which  Goldstone Theorem requires that a  zero energy  
excitation still exists  \cite{boka96}.  
 Extending  the above ideas to non-equilibrium, we will  define the spatial 
   PDF  as
\ba
\left[\fpisp (t)\right]^2&=&\frac{1}{2\pi}\lim_{\omega\rightarrow 0^+}
\int_{-\infty}^{\infty} dq_0 q_0 \rho_L (q_0,\omega,t)
\nonumber\\
&=&
\lim_{\omega\rightarrow 0^+} i\left.\frac{d}{dt}\rho_L (\omega,t,t')
\right\vert_{t=t'}
\label{fpisp}
\ea
where in the second line  we have written $\rho_L$ in the mixed 
 representation. 
 In equilibrium, the above 
 gives  $\fpi$ independent of $\omega$ and then there is no need of taking the 
 $\omega\rightarrow 0^+$ limit \cite{boka96}, meaning that the 
 pion distribution function still behaves as a \mbox{$\delta$-function} 
to this 
 order of approximation. It is not obvious in the least that the same 
 has to  hold out of equilibrium to one-loop, as it is indeed the case 
 (see below). 

Let us 
 check the consistency of the definition (\ref{fpisp}). To leading order, 
 from (\ref{axcurrlo}), we have 
\be
\rho_L^{LO} (\omega_q,t,t')=if(t) f(t') \rho_0 (\omega_q,t,t')
\label{rhollo}
\ee
with $\rho_0$ the  spectral density (\ref{rho}) to leading order. 
Then, using (\ref{rho0norm}), we readily find 
$\fpisp (t)^2_{LO}=f^2 (t)$, 
i.e, the PDF coincides with $f(t)$ to leading order, as 
 it should be. Following the same ideas, we will  introduce  
\ba
\fpisp (t) \fpite (t) &=&\frac{1}{2\pi}\lim_{\omega\rightarrow 0^+}
\int_{-\infty}^{\infty} dq_0  \rho_S (q_0,\omega,t)
\label{fpits}\\
\fpisp (t) \gpit &=&-\frac{i}{2\pi} \lim_{\omega\rightarrow 0^+} 
 \int_{-\infty}^{\infty} dq_0 q_0  \rho_S (q_0,\omega,t)
\label{fpisgpi}
\ea

The function $\fpisp (t)$ is the nonequilibrium extension of the spatial 
 pion decay constant, whereas $\gpit$ vanishes in equilibrium. 
However, according to our previous discussion on the WI, 
the three PDF  we have introduced are not
 independent.  In fact, from their  above definitions and 
 integrating in $q_0$ in   (\ref{wirho}), we find
\ba
\fpisp (t) \gpit&=&\frac{1}{2}\frac{d}{dt} 
\left [\fpite (t) \fpisp (t)\right]
\label{wifpi}
\ea
and then  there are only two independent PDF,  
 also as in equilibrium \cite{pity96}. 

As we did above with $\fpisp (t)$, we can check now the above expressions 
 to leading order. From (\ref{axcurrlo}) we find 
$\fpite (t)_{LO}=\fpisp (t)_{LO}= \ft$ and $\gpit_{LO}=\dot\ft$, so that our 
 definitions are consistent to LO. 



To NLO, we need the axial current to $\Od (\pi^3)$. From (\ref{currents})
 we have
\ba
A_\mu^ a \vxt &=&  -\ft\partial_\mu\pi^a+\delta_{\mu 0}\fdot\pi^a
-\frac{1}{2\ft}\left(\pi^a\partial_\mu \pi^2\right.
\nonumber\\
&-&\left.\pi^2\partial_\mu \pi^a
 - \delta_{\mu 0} \frac{\fdot}{\ft}\pi^2 \pi^a \right)+
 \Od (\pi^5)
\label{axcurrnlo}
\ea

Therefore, according to the chiral power counting explained in previous 
 sections, 
we have two types of contribution to the axial-axial 
 correlator  NLO corrections. 
The first one comes from the NLO correction to the pion 
 propagator, which we have evaluated in section \ref{nloprop}, 
in the product of the $\Od (\pi)$ terms above, whereas 
 the second is the product of the $\Od (\pi)$ with the $\Od (\pi^3)$. 
 These two contributions are represented in Fig.\ref{fig2} by the  diagrams
  a) and b) respectively. As we did  in the evaluation of the NLO
 propagator, we are not  considering  the $\lagf$ contributions 
 (see below).   
After evaluating  the loops, 
 we obtain $\rho_{\mu\nu}$ to NLO. We give here the result 
 for  the spatial-spatial component:
\ba
\lefteqn{\rho_L^{NLO} (\omega_q,t,t')=\rho_L^{LO} (\omega_q,t,t')
 +if(t)f(t')\rho_{tad} (\omega_q,t,t')}
\nonumber\\
 &+&\frac{1}{2}\rho_0 (\omega_q,t,t')\left[\frac{f(t)}{f(t')}G_0 (t')
 +\frac{f(t')}{f(t)}G_0 (t)\right]\label{rholnlo}
\ea
with $\rho_L^{LO}$  in (\ref{rhollo}) 
 and $\rho_{tad}=G^>_{NLO}-G^<_{NLO}-\rho_0$. 
 The $\rho_{tad}$ contribution in the 
 above equation comes  from the diagram a) in Fig.\ref{fig2} and  the 
 rest  from diagram b). 
From  our definitions (\ref{fpisp}) and (\ref{fpits}) we 
  obtain 
\ba
\left[\fpisp (t)\right]^2_{NLO}&=&\left[\fpite (t)\right]^2_{NLO}
=f^2 (t)
\nonumber\\
&-&i\left[G_0(t)-f^2 (t) {\cal H}(t)\right]\label{fpistnlo}
\ea
\ba
\lefteqn{\left[\fpisp (t)\gpit\right]_{NLO}=\ft\dot\ft}
\nonumber\\
&-&i\left[\dot G_0(t) +G_0(t)\frac{\dot\ft}{\ft}+\dot\ft\ft 
{\cal H}(t)\right]
\label{fpisgpinlo}
\ea

with
\be
{\cal H}(t)\equiv\lim_{\omega\rightarrow 0^+}\left.\frac{d}{dt}\rho_{tad} 
(\omega,t,t')\right\vert_{t=t'}=-\frac{G_0(t)}{f^2 (t)}
\label{ht}
\ee
where we have made use of  (\ref{nloposit}), 
 taking, without loss of generality, both $t,t'\in C_1$ and positive (we 
 know the equilibrium answer for the negative real axis). 
 We  realise that $\Delta_1 (t,\omega)$ in  (\ref{nloposit}) 
 does not contribute to ${\cal H}(t)$ and that the whole  NLO 
result for the PDF is independent of $\omega$, so that  there is no 
need of taking the $\omega\rightarrow 0^+$ limit, which 
together with $\fpisp (t)=\fpite (t)$ in (\ref{fpistnlo}) and 
according to our previous discussion, 
we may interpret  as the existence (to this order)   
of undamped NGB in the nonequilibrium  plasma, propagating at the 
 speed of light (see \cite{pity96}). 

In addition, when we substitute (\ref{ht}) in (\ref{fpistnlo}) and 
(\ref{fpisgpinlo}), we see that $\gpit$ to NLO satisfies the WI 
relation (\ref{wifpi}),   which is a  consistency check. 
We finally obtain
\be
\left[\fpisp (t)\right]^2_{NLO}=\left[\fpite (t)\right]^2_{NLO}
=f^2 (t)-2i G_0(t)
\label{fpitnlo}
\ee

This is the one-loop relationship between  the PDF
 and $\ft$ \footnote{Notice that $G_0 (t)$  in (\ref{fpitnlo}) 
depends implicitly on $f(t)$, through  (\ref{loprop}).}, extending the 
equilibrium result (\ref{fpieq}), which we recover (for the general 
 contour $C$) simply  by replacing  (\ref{g0eqdr}) in (\ref{fpitnlo}). 
We want to stress that the PDF are observable and therefore 
(\ref{fpitnlo}) should be  independent of the parametrisation 
 chosen for the pion fields. We have checked it explicitly by calculating 
 with the $\tilde\pi$ fields defined before. Both the propagator and 
 the axial-axial spectral functions change, but  (\ref{fpitnlo}) remains the 
 same  with the same definitions (\ref{fpisp}) and  (\ref{fpits}) and the 
 same $G_0 (t)$. 
 In turn, let us mention that by extending naively the low $T$ results in the 
 LSM (see \cite{boka96}) to nonequilibrium we would have 
 $v^2 (t) =\fpi^2 (t)-iG_0 (t)$  with   $v(t)=\langle \sigma \rangle (t)$. 
We will make use of this  assumption below  to compare our  results 
 with  \cite{ladaco96}.

 Notice also  that it is licit to replace $\ft$ by 
 $\fpit$ for $G_0 (t)$ in  (\ref{fpitnlo}), to the order we are considering 
 here.  Therefore,  using (\ref{fpitnlo}) 
  allows to express all physical non-equilibrium observables 
(like decay rates, masses, etc) obtainable from our model 
 to one loop in ChPT, in 
 terms of the physical $\fpit$, which could be determined, for instance, 
 by analysing non-equilibrium lepton decays $\pi\rightarrow l \nu_l$. 
A very important aspect in this  program 
 is renormalisation. 
 As we said before,  $G_0(t)$  contains in general time-dependent 
 infinities. However, as we are going to see, in certain 
 regimes (for instance, at short times) 
it is still possible to find a finite answer for $G_0 (t)$ in 
dimensional regularisation (as in equilibrium in the chiral 
 limit) so that the counterterms coming from 
 $\lagf$ are not needed. There still could  be finite contributions coming 
 from $\lagf$ but those  are typically of $\Od(10^{-3})$ with 
 respect to those coming from the loops \cite{dogoho92,dole91}, so that 
 our approximation of neglecting $\lagf$ would be justified in that case, 
 which is the one we will consider in the remaining of this work.  
 Explicit 
 knowledge of  $\ft$ would require to solve the hydrodynamic 
 equations for the plasma, or,  equivalently, the Einstein equations 
 for the metric. Alternatively, one can follow the approach of treating 
 $\ft$ as external, choosing for instance different parametrisations 
 compatible with  (\ref{chipoco}). This is left for future investigation.


 For short times, the particular form of $\ft$ is not important and 
 we can  parametrise the  nonequilibrium dynamics 
in terms of a few parameters, which are the values of $\ft$ and 
 its derivatives at $t=0$. This is equivalent to
 describe  the expansion of the Universe in terms of the Hubble parameter, 
 the deceleration parameter and so on. As we discussed in 
 section \ref{model}, this approach is justified for 
times $t<t_{max}$ with 
 $t_{max}\simeq 1/\fpi (0)\simeq$ 2 fm/c, which are indeed not so short, 
 regarding the typical time scales involved in a RHIC (see our comments below 
 and in section \ref{intro}). 
Let us  then start by considering the differential 
equation (\ref{lopropmix})
 with KMS conditions at $t_i$.   
 In addition, as we have discussed in section \ref{model}, 
 we shall impose  that $G_0 (\omega_q,t,t')$ be continuous 
 and differentiable in $t=0$ and  $t'=0$.   
 The solution of  (\ref{lopropmix}) can be written in terms of two arbitrary 
solutions $g_i (t)$ of the  homogeneous equation, satisfying the 
 Wronskian condition $\dot g_1 g_2-g_1\dot g_2=1$. 
In   \cite{sewe85}, the general  solution of (\ref{lopropmix})
 satisfying KMS conditions  was  constructed 
in terms of the $g_i(t)$ functions, 
  for $t$ and $t'$ in  the contour $C$ (as we have said before,   
 it is enough for us to take   $t,t'\in C_1$). We remark 
 that the solution of the homogeneous equation is only  
 known explicitly for a some special choices of    $m^2(t)$ 
\cite{sewe85,bodeho95}.

Therefore, we expand both  $\ft$ and the solution of the 
 homogeneous equation  near  $t=0$. 
Both the differential equation and the Wronskian condition 
 reduce then to algebraic relations between the first coefficients in the 
 expansion. With the boundary, continuity and differentiability 
 conditions discussed above, the solution is uniquely defined and 
we find to  lowest order
\be
G_0(\omega_q,t,t')=-\frac{i}{2\omega_q}
\coth\left[\frac{\beta_i\omega_q}{2}\right]\left[1-m^2t^2 + 
\Od (m^4t^4)\right]
\label{g0st}
\ee
with $m^2\equiv m^2 (0)=-\ddot f(0)/f(0)$. For $m^2<0$ 
we see the onset of unstable modes, 
 making the pion correlation function grow with time. 
 We also observe  that for short times, the time dependence of the mixed 
 propagator factorises. Therefore, the momentum dependence 
 is the same as in equilibrium and then we can  integrate it in 
 dimensional regularisation, using  
(\ref{g0eqdr}), to  get a finite answer. As we said before, 
 in this regime we are able to avoid renormalisation. 
 Then, from (\ref{fpitnlo}) and  
 (\ref{g0st}) we get
\be
\fpitsq=f^2\left[a(T_i)+2Ht-c(T_i)t^2+\Od(|m|^3t^3)\right]
\ee
where $f=f(0)$,  $H=\dot f(0)/f(0)$ ($H$ and $m$ are $\Od (p)$ and 
 play the role of the Hubble and 
 deceleration parameters in the expansion, respectively)   and 
\ba
a (T_i)&=&1-\frac{T_i^2}{T_c^2} \nonumber\\
c(T_i)&=&m^2 a(T_i)-H^2
\label{nlocoeff}
\ea
where  $T_c\equiv \sqrt{6}\fpi \simeq 228$ MeV.
 Therefore, (\ref{nlocoeff}) provides the nonequilibrium 
relationship 
between 
 the LO  Taylor coefficients, independent of $T_i$, 
 appearing in the lagrangian and the physical $T_i$-dependent 
NLO ones, extending the equilibrium formula (\ref{fpieq}) when $f$ is the 
only lagrangian parameter.  The coefficients of order zero and two in the 
 Taylor expansion of $\fpit$ are corrected by the same one-loop 
$T_i$-dependent factor, whereas the first order coefficient $H$  does not 
 acquire NLO corrections. We would like to remark that the corrections 
 coming from $\lagf$ that we have neglected are, in addition,  
 independent of $T_i$, since they only contribute at tree level to NLO.

We will proceed now to discuss some physical effects related to 
the behaviour of $\fpit$. 
 For that purpose, and based upon (\ref{fpieq}), let us  
 define a time-dependent effective temperature $T(t)$ as
\be
\fpitsq =f^2\left[ 1-\frac{T^2(t)}{T_c^2}\right]\quad ,
\label{Tt}
\ee
i.e, we parametrise  the nonequilibrium effects of $\fpi (t)$ in $T(t)$. 
Let us define  in addition  the critical 
 time $t_c$, as   $\fpi (t_c)=0$ ($T(t_c)=T_c$) and  the freezing 
 time $T(t_f)=0$, $\fpi (t_f)=f$. These are the relevant time scales 
 when the system is heating or cooling down, respectively. 
The sign of $H$ determines whether the 
 plasma is initially heated ($H<0$) or cooled down ($H>0$). 
However, it is the sign of $c(T_i)$ in (\ref{nlocoeff}) what determines 
 the behaviour at longer times. Thus, we will follow the short-time 
 evolution of $\fpit$ until it reaches either $t_c$ or $t_f$, imposing 
 that $0<T(t)<T_c$. We are following a similar approach as that of  
equilibrium  when one 
 extrapolates  (\ref{fpieq})  until $T_c$.

%
%
%
%
%
%
%

In order to estimate  the above time scales, let us take 
$H^2\simeq|m^2|$  and  retain only 
 the leading order in $x\equiv T_i^2/T_c^2$, consistently 
 with the chiral expansion. Then, we have the following cases:
\begin{enumerate}

\item  $H>0$: Cooling down until $\tti_f\simeq x/2$.

\item  \mbox{$H<0$}, \mbox{$m^2>0$}: 
 Heating until \mbox{$\tti_c\simeq (1-3x/4)/2$}.

\item  $H<0$, $m^2<0$: Heating until 
\mbox{$\tti_m\simeq (1+x/2)/2$} 
 and then cooling down until \mbox{$\tti_f\simeq 1+x$}, 
passing through the 
 equilibrium time 
\mbox{$\tti_{eq}=2\tti_m$} such that $\fpi (t_{eq})=\fpi (0)$.

\item  $H=0$, $m^2>0$: 
 Heating until  $t_c^2 m^2=1$, independent 
 of $T_i$.

\item  \mbox{$H=0$}, \mbox{$m^2<0$}: Cooling   until 
\mbox{$t_f^2 m^2\simeq -x(1+x)$}
\end{enumerate}
where $\tti=t|H|$. 
 We observe that the effect of the unstable modes ($m^2<0$) is always 
to cool down the system.  The freezing time $t_f$ for $H<0$  is 
  much longer (according to our power counting)  
 than that for $H>0$,  which is $\Od(x/H)$. Indeed, the $\Od (1/H)$
 time scales   are much 
 longer than those to which our short-time approximation 
 remains valid and they have to be understood as
 estimates, similarly to the equilibrium case, where one estimates  
 the critical temperature, 
even though the low $T$ approach breaks down for $T\simeq T_c$. 
 Taking  typical  values  
$T_i\simeq |H|\simeq |m|\simeq 100$ MeV ($x\simeq 0.16$) we get 
 $t_f\simeq$ 0.2 fm/c for cases 1 and 5, $t_c\simeq 2$ fm/c for cases 2 and 
 4 and $t_f \simeq$ 2.3 fm/c for case 3.

On the other hand, evaluating  the above time scales for initial 
temperatures closer to $T_c$, by 
 expanding now for small $a(T_i)$ in  (\ref{nlocoeff}), we get
 $\tti_f\simeq \sqrt 2 -1$ for case 1, 
 $\tti_c\simeq a(T_i)/2$ for case 2,  $\tti_m\simeq 1$, 
$\tti_f\simeq 1+\sqrt 2$ for case 3 and  $t_f^2 m^2= -x/a(T_i)$ for case 5. 
 We can compare  with  
$v (t)$ in \cite{ladaco96} (see our comment above), 
 using their
 initial  values $T_i\simeq$ 200 MeV and $|H|\simeq$ 400 MeV (which are
 clearly too high for our low-energy approach). 
We  see that our estimates for the  time evolution duration are of the 
 same order as those in \cite{ladaco96}, although ours are somewhat lower.  
 An important remark 
is that in typical LSM simulations like \cite{ladaco96}, 
$v (t)$ reaches a stationary value, about which it 
 oscillates, the relevant time scale being the time it takes the system to
 reach such  value. 
 It is clear that we cannot predict that type of behaviour only 
 within  our short-time approach, quadratic in time, but only estimate the 
 time scales involved \footnote{Similarly as to why
ChPT cannot see the phase transition 
(see comments in section \ref{intro})}, i.e, 
 we cannot see the system thermalise  in that sense.



\section{Conclusions and Outlook}
\label{conc} 
 
We have extended   chiral lagrangians and ChPT  
 out of thermal equilibrium. 
The chiral power counting requires  
 all time derivatives to be $\Od (p)$ and  to lowest order
 our model is a 
 NLSM with $f\rightarrow \ft$. This model 
 accommodates unstable pion modes and 
 corresponds to a spatially flat RW metric in conformal time 
 with $a(t)=f(t)/f(0)$ and minimal coupling. 
 For a  non-minimal coupling, the chiral symmetry would be  broken 
 and conformal invariance  restored for $\xi=1/6$.

We have applied our  model to study 
 the  time dependent PDF, which extend the equilibrium 
 pion decay constants. We find that in general there are two 
 independent PDF, as in   equilibrium. To one loop in ChPT such  functions 
 coincide and we have given their expression in terms 
 of the equal-time correlation function, analysing  the lowest order 
short-time coefficients of $\fpit$. We have studied the relevant time 
 scales and their dependence with $T_i$, paying special attention to the 
 role of the unstable modes in  cooling down the plasma. 
 In this work we have treated  nonequilibrium renormalisation 
  by remaining  within the short-time expansion, where the equal-time 
 correlator  turns 
 out to be free from UV divergences in the chiral limit in dimensional 
 regularisation and hence   it is 
 reasonable to neglect the contributions coming from  $\lagf$ in our 
 analysis.  
 Nonetheless,  we have seen that, thanks to the analogy with 
curved space-time,
 we have a well-defined procedure to  construct $\lagf$ so as to absorb 
 all the loop divergences, which in general will be time-dependent, and 
 find a finite answer for the observables.  

Among the aspects we would like to study further is to include 
 the nonequilibrium $\lagf$ tree contributions, so that we can renormalise 
 $\fpi (t)$  $\forall t$ and analyse the long time behaviour properly. 
An interesting remark is that for long times with 
 unstable modes of negative mass squared $m^2$, 
only the $k^2<|m^2|$ modes are important, so that there is 
 an effective ultraviolet cutoff and we can forget  about 
 UV divergences in that case 
as well. Another point we leave for future investigation 
is the behaviour of the two-point correlation function 
at different space points 
 for different choices of $\ft$, which would allow us to investigate 
 the formation 
 of regions of unstable vacua (DCC) in our model. 
Other applications and extensions of our approach, 
to be explored in the future,   include  photon production in the pion 
sector (by gauging the theory and including  $\pi^0$ decay photon production 
 through  the anomalous Wess-Zumino-Witten  term),  the quark condensate 
 time dependence (by considering the mass explicit symmetry-breaking terms
 we have neglected here) and the $N_f=3$ case. Finally, we hope that by 
 implementing resummation methods like large $N$ in our model, we 
 may be able to study  the long-time thermalisation process.

\section*{ACKNOWLEDGMENTS}
This work would not have been possible without the help, through multiple 
 and useful discussions,  of   V.Gal\'an-Gonz\'alez, T.Evans and  R.Rivers, 
 from the Imperial College Theory Group, to whom I have to thank  their 
  kind  hospitality. I am also grateful to  A.Dobado and R.F.Alvarez-Estrada
 for providing useful references and comments, as well as to the organisers
 and participants  of TFT'98 for a very stimulating workshop.  
My stay at Imperial College was funded by a
 fellowship of the Spanish Ministry of Education. I also 
 acknowledge financial support  from  CICYT, Spain, project AEN96-1634.

\end{document}